\pdfoutput=1
\documentclass[12pt]{article}
\usepackage[utf8x]{inputenc}
\usepackage[english]{babel}
\usepackage{amsfonts,amsmath,amssymb}
\usepackage{mathtools}
\usepackage{enumerate}
\usepackage{booktabs}
\usepackage{cite}
\usepackage{pgfplots}
 \pgfplotsset{compat=1.9}
\usepackage{caption}
\usepackage{subcaption}
\usepackage[breaklinks,colorlinks=true,urlcolor=RoyalBlue,linkcolor=blue,
citecolor=blue]{hyperref}
\catcode`@=11
\allowdisplaybreaks

\newcommand{\pa}{\partial}
\newcommand{\diff}{\mathrm{d}}
\newcommand{\nn}{\nonumber}
\newcommand{\Ord}{\mathcal{O}}

\newcommand{\DelU}{\Delta U}

\makeatletter

\@addtoreset{equation}{section}
\makeatother

\def\href#1#2{#2}

\begin{document}
\begin{titlepage}
\begin{flushright}
{UWTHPH-2017-2}
\end{flushright}

\begin{center}

\hfill 
\vskip 1.2in

\textbf{\LARGE Tunnelling in Dante's Inferno}\\[18mm] 

{Kazuyuki Furuuchi${\,}^a$ and Marcus Sperling${\,}^{b,(a,c)}$}
\vskip8mm
${}^a\,${\sl Manipal Centre for Natural Sciences, Manipal University}\\
{\sl Dr.T.M.A. Pai Planetarium Building}\\ 
{\sl Madhav Nagar, Manipal, Karnataka 576104, India}\\[2mm]
${}^b\,${\sl Fakult\"{a}t f\"{u}r Physik,
Universit\"{a}t Wien}\\ 
{\sl Boltzmanngasse 5, A-1090 Wien, Austria}\\[2mm]
${}^c\,${\sl Institut f\"{u}r Theoretische Physik,
Leibniz Universit\"{a}t Hannover}\\ 
{\sl Appelstraße 2, 30167 Hannover, Germany}

\vskip10mm 
\end{center}
\begin{abstract}
We study quantum tunnelling in Dante's Inferno model of large field inflation.
Such a tunnelling process, which will terminate inflation, becomes problematic 
if the tunnelling rate is rapid compared to the Hubble time scale at the time 
of inflation.
Consequently, we constrain the parameter space of Dante's Inferno model by 
demanding a suppressed tunnelling rate during inflation.
The constraints are derived and explicit numerical bounds are provided for 
representative examples.
Our considerations are at the level of an effective field theory;
hence, the presented constraints have to hold regardless of any UV completion.
\end{abstract}

\end{titlepage}

\tableofcontents

\section{Introduction}\label{sec:Intro}

The slow-roll inflation paradigm has been phenomenologically successful, 
initially solving the naturalness issues in Big Bang Cosmology,
and later explaining the primordial density perturbations.
General predictions of slow-roll inflation on primordial density perturbations
agree very well with recent Cosmic Microwave Background (CMB) observations.

Nevertheless, slow-roll inflation has its own naturalness issue.
Protecting the flatness of the inflaton potential against quantum corrections
has been a long-standing challenge.
This issue is particularly severe in large field inflation models in which 
the inflaton enjoys super-Planckian field excursion.

A standard approach to explain the flatness of a potential in an effective 
field theory is imposing a symmetry.
For example, natural inflation \cite{Freese:1990rb} assumes a continuous shift
of an axion field as an approximate symmetry.
The comparison of this model with CMB data requires a super-Planckian axion
decay constant.
Naively, this indicates that the symmetry must be respected at the Planck scale.
However, there are strong indications that continuous global symmetries are not 
respected in a quantum theory of gravity (see \cite{Banks:2010zn} for a recent 
discussion together with a review of earlier studies).

An approach to circumvent this problem was proposed under the name of 
extra-natural inflation \cite{ArkaniHamed:2003wu}.
This model realises a super-Planckian axion decay constant in four dimensions
by means of an effective gauge field theory in higher dimensions.
The super-Planckian axion decay constant is achieved at the expense of a very 
small gauge coupling.
However, it was immediately noticed that this model is difficult to realise in 
string theory \cite{ArkaniHamed:2003wu,Banks:2003sx}.
The lasting difficulty in realising extra-natural inflation in string theory
led to the \emph{Weak Gravity Conjecture} \cite{ArkaniHamed:2006dz},
which limits the relative weakness of gauge forces compared to the gravitational 
force.
This conjecture may eventually forbid a super-Planckian axion decay constant
in effective field theories which can consistently couple to gravity,
though several logical steps need to be examined in more detail.

If a super-Planckian axion decay constant is forbidden in effective field 
theories which are consistently coupled to gravity, then a new major 
obstacle for the realisation of large field inflation via natural 
inflation arises.
However, a possible way out may be axion monodromy 
inflation \cite{Silverstein:2008sg,McAllister:2008hb,%
Kaloper:2008fb,Berg:2009tg,Kaloper:2011jz}.
In this class of models, the axion decay constant is sub-Planckian,
but the axion couples to an additional degree of freedom, which we call 
\emph{winding number direction} below.
An effective super-Planckian excursion of the inflaton is achieved by going 
through the axion direction multiple times, with a shift in the winding number 
direction for each round.
This appears to be a promising avenue for realising large field inflation.
Nevertheless, the validity of axion monodromy inflation should be examined 
further, both at the level of an effective field theory as well as at the level 
of an UV completion.
In particular, it has been pointed out that quantum tunnelling through the 
potential roughly in the winding number direction may terminate inflation 
before it lasts long enough for solving the naturalness issues in Big Bang 
Cosmology \cite{Kaloper:2011jz}.
It turned out that the tunnelling rate is highly model dependent.
Tunnelling in related models has
subsequently been studied in 
\cite{Franco:2014hsa,Harigaya:2014rga,Blumenhagen:2015kja,%
Ibanez:2015fcv,Hebecker:2015zss,Brown:2016nqt,McAllister:2016vzi}.

In this article, we study tunnelling in an axion monodromy model, 
namely \emph{Dante's Inferno model} \cite{Berg:2009tg}.
We limit our study to the level of an effective field theory.

Besides phenomenological interests in this promising model,
there is an attractive technical feature: The potential wall
orthogonal to the inflaton direction is explicitly given.
This allows us to apply a standard calculation \`{a} la 
Coleman \cite{Coleman:1977py} in order to estimate the tunnelling rate.
In particular, one can estimate the tension of the surface of the 
bubble, through which the false ``vacuum" decays\footnote{%
Precisely speaking, during inflation
the state is not in a local minimum of the potential,
but slowly rolling in $\phi$-direction.
This point has been investigated in \cite{Brown:2016nqt}.
In this article, we will loosely use the term (false) ``vacuum"
for such configurations, because we will be mainly dealing with
 slices of constant $\phi$ of the potential
in which the state is in a local minimum.}.
This is in contrast to other axion monodromy models for which
the tension of the wall is treated as an input from a UV theory
\cite{Kaloper:2011jz,Franco:2014hsa,%
Blumenhagen:2015kja,Ibanez:2015fcv,%
Hebecker:2015zss,Brown:2016nqt,McAllister:2016vzi}.

We constrain the parameter space of Dante's Inferno model by requiring a 
suppressed tunnelling rate during inflation.
In particular, we will show that in some regions of the parameter space,
the suppression of the tunnelling process yields a new constraint.
This constraint comes purely at the level of an effective field theory;
hence, regardless of the UV completion of the theory, the constraint has to 
hold.
For a fixed ratio $\Lambda/f_1$, where $\Lambda$ is the parameter controlling 
the height of the sinusoidal potential and $f_1$ is the smaller axion decay 
constant in Dante's Inferno model, the condition that tunnelling is suppressed 
introduces a lower bound on $f_1$ in such a parameter region. 
We demonstrate this observation by providing explicit numerical bounds
in a couple of representative examples.

The outline of this article is as follows: Dante's Inferno model is briefly 
reviewed in Sec.\ \ref{sec:DI}. 
Thereafter, we discuss quantum tunnelling and suppression thereof in 
Sec.\ \ref{sec:TDI}. 
We exemplify these considerations for the choice of a monomial inflaton 
potential in Sec.\ \ref{sec:Ex}. 
Lastly, Sec.\ \ref{sec:discussions} concludes. 
Three appendices provide the necessary background and details for choosing 
constant field values during inflation, the bounce 
solution, and the thin-wall approximation.
%
%
\section{Dante's Inferno model}\label{sec:DI}

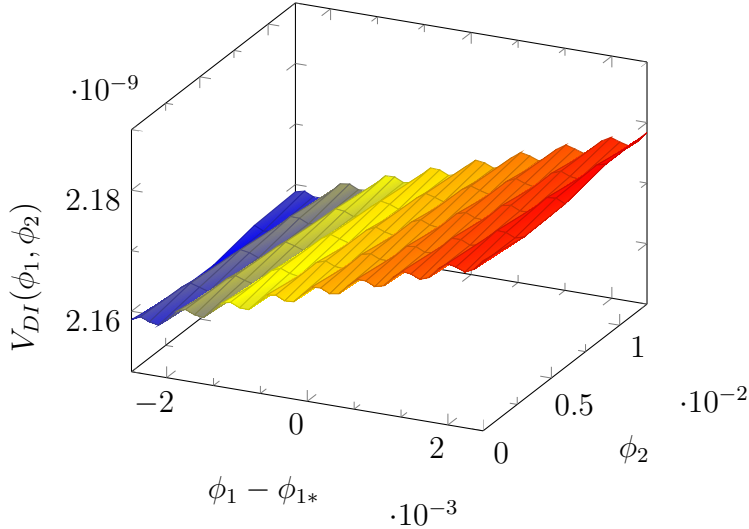
\begin{figure}[h]
\centering
\begin{tikzpicture}
\begin{axis}[
xmin=-0.25e-02, xmax=0.25e-02,
ymin=0, ymax=120e-04,
 zmin=2.15e-09, zmax=2.19e-09,
 minor x tick num=2,
minor y tick num=1,
minor z tick num=1,
xlabel = {$\phi_1-\phi_{1 \ast}$},
ylabel = {$\phi_2$},
zlabel = {$V_{DI}(\phi_1,\phi_2)$}]
\addplot3[width=1\textwidth,surf,shader=faceted interp,
samples = 40, 
samples y = 5, 
domain=-0.25e-02:0.25e-02,
y domain = 0:120e-04] 
{(1.9698e-10)*(x+0.55)*20 + (1e-12)*( 1-cos(deg(((x+0.55)-y/20)*(1e04)) ) )};
\end{axis}
\end{tikzpicture}
\caption{The potential $V_{DI}(\phi_1,\phi_2)$ of Dante's Inferno model for the 
exemplary parameters $\Lambda=10^{-3}$, $f_1=10^{-4}$, and $f_2 =20 \cdot f_1$, 
together with a monomial inflaton potential, c.f.\ Sec.\ \ref{sec:Ex} with 
parameters $p=1$ and $N=60$. The reference point $\phi_{1 \ast}$ is defined 
around \eqref{sing}. The boundaries of the $\phi_2$-direction in the plot 
should 
be identified. Going around in $\phi_2$-direction through the valley of the 
potential results in a shift in $\phi_1$-direction.}
\label{fig:DIpot}
\end{figure}

In this section, we review Dante's Inferno model \cite{Berg:2009tg} 
and fix our notation. The dynamics of the model are governed by the following 
action:
\begin{subequations}
\begin{equation}
S_{DI} =
\int d^4 x
\sqrt{-g}
\Biggl[
\frac{1}{2} \pa_\mu \phi_1 \pa^\mu \phi_1
+
\frac{1}{2} \pa_\mu \phi_2 \pa^\mu \phi_2
- 
V_{DI}(\phi_1,\phi_2)
\Biggr] ,
\label{SDI}
\end{equation}
where the scalar potential is given by
\begin{align}
V_{DI}(\phi_1,\phi_2)
=
V_1(\phi_1)
+ 
{\Lambda^4}
\left( 1 - \cos \left(\frac{\phi_1}{f_1} - \frac{\phi_2}{f_2} \right)
\right).
\label{VDI}
\end{align}
\end{subequations}
Fig.\ \ref{fig:DIpot} displays the behaviour of the potential 
$V_{DI}(\phi_1,\phi_2)$ for some parameter values.
%
%
%
It is convenient to perform the following rotation in the field space:
\begin{subequations}
\begin{equation}
\left(
\begin{array}{c}
\chi \\
\phi
\end{array}
\right)
=
\left(
\begin{array}{cc}
\cos \gamma & - \sin \gamma\\
\sin \gamma & \cos \gamma
\end{array}
\right)
\left(
\begin{array}{c}
  \phi_1\\
	\phi_2
\end{array}
\right) ,
\label{rot}
\end{equation}
where
\begin{equation}
\sin \gamma \coloneqq \frac{f_1}{\sqrt{f_1^2 + f_2^2}},
\qquad
\cos \gamma \coloneqq \frac{f_2}{\sqrt{f_1^2 + f_2^2}}  .
\label{angle}
\end{equation}
\end{subequations}
In terms of the rotated fields, the potential \eqref{VDI} becomes
\begin{subequations}
\begin{equation}
V_{DI}(\chi,\phi)
=
V_1 (\chi \cos \gamma + \phi \sin \gamma)
+ 
\Lambda^4 
\left(
1 - \cos \frac{\chi}{f}
\right),
\label{tildepot}
\end{equation}
where
\begin{equation}
f \coloneqq \frac{f_1 f_2}{\sqrt{f_1^2 + f_2^2}} .
\label{f}
\end{equation}
\end{subequations}
According to \cite{Berg:2009tg}, the following two conditions are required
for Dante's Inferno model:
\begin{subequations}
\begin{align}
&2\pi f_1 \ll 2\pi f_2 \lesssim M_P \; ,
\label{DI1}\\
&\frac{\Lambda^4}{f} \gg V_1'\; .
\label{DI2}
\end{align}
\end{subequations}
Here, $M_P \coloneqq (8\pi G)^{-1/2}$ is the reduced Planck mass
with $G$ being Newton's constant.
For later convenience, we rewrite condition \eqref{DI2} as
\begin{equation}
s \coloneqq \frac{\Lambda^4}{f V_1'} \gg 1.
\label{slopes}
\end{equation}
The last inequality in \eqref{DI1} is expected to follow from the Weak Gravity 
Conjecture as we reviewed in the Introduction, which we assume in this article.
Next, condition \eqref{DI1} implies
\begin{equation}
f \simeq f_1\, , 
\qquad
\cos \gamma \simeq 1 \, , 
\qquad 
\sin \gamma \simeq \frac{f_1}{f_2} 
\ll 1 \;.
\label{fapprox}
\end{equation}
Now, let us take a closer look on the origin of condition \eqref{DI1}.
In large field inflation, we typically have $\phi_\ast \gtrsim 10 M_P$,
where the suffix $\ast$ indicates that it is the value when the pivot scale
exited the horizon (see \eqref{phiast} in Sec.\ \ref{sec:Ex}
for values of $\phi_\ast$ in explicit examples).
This constrains $\sin \gamma$ via
\begin{equation}
\sin \gamma 
=
\frac{\phi_{1 \ast}}{\phi_\ast} .
\label{sing}
\end{equation}
For the effective field theory description of the potential $V_1(\phi_1)$ to be 
valid, a natural expectation is that $\phi_{1 \ast}$ is bounded from above by 
the reduced Planck scale $M_P$.
This assumption together with \eqref{sing} implies that $\sin \gamma \lesssim 
0.1$
whenever $\phi_\ast \gtrsim 10 M_P$.
Note that the effective description may break down
at a much smaller energy scale, such that the value of $\phi_{1 \ast}$ 
decreases accordingly.
For example, for a moderate model assumption $\phi_{1 \ast} \lesssim 10^{-1} 
M_P$, then \eqref{sing} imposes $\sin \gamma \lesssim 10^{-2}$.

Next, let us examine condition \eqref{DI2}, which implies that the field 
$\chi$ first settles down to the local minimum in a slice of constant $\phi$ 
before the field $\phi$, which plays the role of inflaton in Dante's Inferno 
model, starts to slow-roll.
Then, from \eqref{tildepot} the inflaton potential $V_I(\phi)$ is given by
\begin{equation}
V_I(\phi) = V_1 (\phi \sin \gamma) \; .
\label{VI}
\end{equation}
We refer to Fig.\ \ref{fig:DIpot} to illustrate that the inflaton rolls along 
the bottom of the valley. As one observes, there seem to be numerous 
valleys in the potential, but all of them are connected
by the periodic identification in $\phi_2$-direction.
As the inflaton rolls along the valley
one period in $\phi_2$-direction,
the bottom of the valley is shifted in $\phi_1$-direction.
While the axion decay constant $f_2$ is sub-Planckian as in \eqref{DI1},
super-Planckian inflaton excursion can be achieved
by going round in $\phi_2$-direction several times.

However, the slow-roll inflation may terminate if quantum tunnelling through 
the 
wall of the valley happens.
Requiring that the tunnelling rate is sufficiently small compared to the 
Hubble time scale during inflation may impose further constraints on the 
parameter space of Dante's Inferno model.
We will explore the consequences of this requirement in the next section.
%
\subsection{Dante's Inferno model from higher dimensional gauge theories}
\label{subsec:HDG}
Dante's Inferno model can be obtained from higher dimensional gauge theories.
In this circumstance there is an additional constraint on the 
parameters \cite{Furuuchi:2014cwa,Furuuchi:2015jfj}, which reads 
\begin{equation}
	\Lambda^4 
	\simeq
	\frac{3 c}{\pi^2 (2\pi L_5)^4}  \; ,
	\label{HDGL}
\end{equation}
where the natural value of $c$ is $\Ord(1)$.
The axion decay constants $f_1$ and $f_2$ are given as
\begin{equation}
	f_1 = \frac{1}{g_1 (2\pi L_5)} \; , \qquad
	f_2 = \frac{1}{g_2 (2\pi L_5)} \; ,
	\label{fs}
\end{equation}
where $g_1$ and $g_2$ are the gauge couplings in four-dimension.
From \eqref{HDGL} and \eqref{fs}, and assuming that the perturbative 
approximation is valid, i.e.\ $g_1 \lesssim 1$, we obtain
\begin{equation}
	\Lambda \lesssim f \; .
	\label{HDGLf}
\end{equation}
%
%
\section{Tunnelling in Dante's Inferno model}
\label{sec:TDI}
%
It is well-known that a quantum field theory with two local minima, $\psi_\pm$, 
of the potential has two classically stable equilibrium states. However, 
assuming that $\psi_-$ is the unique state with lowest energy, the state 
$\psi_+$ is rendered unstable quantum mechanically, because of a non-vanishing 
tunnelling probability through the potential barrier into the so-called true 
vacuum state $\psi_-$.
The decay of a false vacuum $\psi_+$ proceeds by nucleation of bubbles, inside 
which the true vacuum\footnote{Below will study tunnelling between a false 
vacuum and another false vacuum with lower energy, which can be analysed 
without introducing new ingredients.} resides.
The tunnelling rate per volume $\Gamma\slash \mathrm{Vol}$ between true
and false vacua, as discussed in \cite{Coleman:1977py,Callan:1977pt}, 
can be parametrised by two quantities $A$ and $B$ (in leading order) via 
\begin{equation}
 \Gamma \slash \mathrm{Vol}= A\ e^{-B \slash \hbar}\ \left[1+ \Ord{(\hbar)} 
\right] \; . 
\label{def:rate}
\end{equation}
While the details of the coefficient $A$ are somewhat complicated, it is 
possible to provide a closed expression for $B$ solely from the semi-classical 
treatment. The relevant solution has been referred to as \emph{bounce} and is 
reviewed in App.\ \ref{sec:Parke}.
From \eqref{def:rate} it is apparent that the tunnelling process is suppressed 
provided $B \gg \hbar$ and the pre-factor $A$ is well-behaved.
A dimensional analysis of the pre-factor reveals $A \sim M^4$, 
where $M$ is a relevant mass scale in the model.
(We refer, for example, to \cite{Baacke:2003uw,Dunne:2005rt} for numerical 
calculations of the coefficients in the case of a simple scalar field theory.)
This estimate may be off by a few orders, but the error will still be small 
compared to the exponential suppression factor $e^{-B \slash \hbar}$.
However, since $B$ is positive, there may be scenarios in which the tunnelling 
is not exponentially suppressed, i.e. $e^{-B \slash \hbar}\sim \Ord(1)$. For 
instance in inflation models, if $A$ induces a rapid rate compared to 
the Hubble time scale during inflation, then the tunnelling becomes 
potentially dangerous as it might terminate inflation too early. More 
precisely, 
this happens if $A \gtrsim H^4$, where $H$ is the Hubble expansion rate at the 
time of inflation.
Consequently, two cases arise: 
\begin{itemize}
 \item On the one hand, if all relevant scales in the 
model are smaller than $H$, the tunnelling rate is irrelevant during inflation, 
regardless of the precise order of $B$.
  \item If, on the other hand, we 
\emph{assume} that all relevant scales in Dante's Inferno model satisfy 
$\Lambda, f_1, f_2 \gtrsim H$ then one has to carefully verify which subsequent 
parameter regions are protected from an unsuppressed tunnelling rate.
\end{itemize}
It is therefore the objective of this article to 
analyse the exponent $B$ together with the condition $B\gg1$ for Dante's 
inferno model for inflation in the regime $\Lambda, f_1, f_2 \gtrsim H$. As 
customary, we set $\hbar \equiv 1$ for the rest of this article.
Obtaining a viable parameter region in Dante's Inferno model then means that 
one has to avoid scenarios in which the tunnelling in $\chi$-direction 
is unsuppressed.
In those cases, one can investigate the dynamics of the field $\chi$, while 
regarding the value of $\phi$ as being fixed 
in time\footnote{A path with 
varying $\phi$ gives a larger action and is irrelevant for the 
estimation of the tunnelling rate.}. We refer to App.\ \ref{sec:constant_field} 
for a discussion of the effects of a time-dependent $\phi$.
Since we are interested in the tunnelling rate during the slow-roll inflation, 
we choose the value of the inflaton when the pivot scale exited the horizon, 
$\phi = \phi_\ast$, as a reference point.
(We comment briefly on other values of $\phi$ at the end of 
Sec.\ \ref{subsec:flat}.)
Then, from \eqref{tildepot} the potential $V(\chi)$ for the field $\chi$
becomes
\begin{equation}
V(\chi) \coloneqq
V_1(\chi \cos \gamma + \phi_\ast \sin \gamma) + 
\Lambda^4 
\left(
1 - \cos \frac{\chi}{f}
\right).
\label{Vchi}
\end{equation}
We first estimate the tunnelling rate including the effects of gravity in order 
to understand when we can neglect the gravitational back-reactions.
Following \cite{Coleman:1980aw}, the Euclidean action of a scalar 
field $\chi$ coupled to Einstein gravity reads
\begin{equation}
S_E
=
\int d^4 x
\sqrt{g}
\left[
\frac{1}{2}
g^{\mu\nu}
\pa_\mu \chi
\pa_\nu \chi
+
V(\chi)
-
\frac{1}{16\pi G}
R
\right].
\label{SE}
\end{equation}
To estimate the gravitational back-reaction, we employ an $O(4)$-symmetric 
ansatz.
There are few limitations of such an ansatz:
Firstly, inflation with (almost) flat spatial space, which is supported by 
observations, does not respect $O(4)$ symmetry\footnote{%
See \cite{Brown:2016nqt} for a recent study
of a non-$O(4)$-symmetric bounce solution 
without gravitational back-reactions.}.
Secondly, there is no proof that the $O(4)$-symmetric bounce gives the least 
action among all bounce solutions.
We will not try to fully justify the use of an $O(4)$-symmetric bounce in this 
article. Nevertheless, since the space is empty during inflation, and we will 
be 
interested in processes which proceed fast compared to the Hubble expansion 
rate, we hope that the first point may not be so crucial.
For the second point, we expect that even if there exists a non-$O(4)$-symmetric
bounce, with smaller action than the $O(4)$-symmetric bounce,
the $O(4)$-symmetric bounce provides at least the
lower bound for the tunnelling rate.
Moreover, we may expect that the difference between the constraints on the 
parameter space of Dante's Inferno model from the non-$O(4)$-symmetric bounce 
do not differ qualitatively from those of the $O(4)$-symmetric bounce.

Assuming $O(4)$ symmetry, the metric takes the form
\begin{equation}
ds^2 =
d \xi^2
+
a^2(\xi)
d\Omega^2,
\label{metric}
\end{equation}
where $d\Omega^2$ is the canonical metric of the unit $S^3$.
Moreover, the $O(4)$ symmetry restricts the field $\chi$ to be a function 
of the radial coordinate $\xi$ only.
Thus, for $O(4)$-symmetric solutions, the Euclidean action \eqref{SE} becomes
\begin{equation}
S_E
=
2\pi^2
\int d \xi
\left[
a^3
\left(
\frac{1}{2}
\left(\frac{d\chi}{d\xi}\right)^2
+
V(\chi)
\right)
-
\frac{3}{16\pi G}
a \left( \left(\frac{da}{d\xi}\right)^2 + 1 \right)
\right].
\label{SEa}
\end{equation}
We have dropped a surface term, which is irrelevant, because we consider the 
difference of actions with the same boundary conditions \cite{Coleman:1980aw}.
It is convenient to rescale the variables as follows:
\begin{equation}
\psi \coloneqq \frac{\chi}{f} \; , 
\qquad
\rho \coloneqq f a \; ,
\qquad
\zeta \coloneqq f \xi \; .
\label{rescale}
\end{equation}
Then \eqref{SEa} becomes
\begin{subequations}
\label{SE_explicit}
\begin{equation}
S_E
=
2 \pi^2
\int d\zeta
\left[
\rho^3
\left(
\frac{1}{2}
\dot{\psi}^2
+
U (\psi)
\right)
-
\frac{3}{\kappa}
\rho
\left(
\dot{\rho}^2 + 1 
\right)
\right] \; ,
\label{SErho}
\end{equation}
where
\begin{align}
\kappa
&\coloneqq
8\pi G f^2 \; ,
\label{kappa} \\
U(\psi)
&\coloneqq
U_0(\psi)
+
U_1(\psi) \; ,
\label{Upsi} \\
U_0(\psi)
&\coloneqq
\lambda^4
\left(
1 - \cos \psi
\right),
\label{U0} \\
\lambda
&\coloneqq
\frac{\Lambda}{f} \; ,
\label{le} \\
U_1(\psi)
&\coloneqq
\frac{1}{f^4} V_1 (f \psi \cos \gamma + \phi_\ast \sin \gamma) \; .
\label{U1}
\end{align}
\end{subequations}
The Euclidean equations of motion are given as
\begin{subequations}
\begin{align}
\ddot{\psi}
+
3
\frac{\dot{\rho}}{\rho}
\dot{\psi}
&=
U'(\psi) ,
\label{psieq}\\
\dot{\rho}^2 - 1
&=
\frac{\kappa}{3}
\rho^2
\left(
\frac{1}{2}
\dot{\psi}^2
-
U(\psi)
\right),
\label{rhoeq}
\end{align}
\end{subequations}
where \eqref{rhoeq} is the Friedmann equation.
The bounce action $B$ reads
\begin{equation}
B \coloneqq
S_E[\psi_B] - S_E[\psi_+] \; ,
\label{B}
\end{equation}
where $\psi_B$ is the bounce solution, and $\psi_+$ is the value of the field 
$\psi$ at the false vacuum we start with, $\psi_+ =0$ in our case.

Similarly to \cite{Coleman:1977py}, we evaluate the bounce 
action \eqref{B} in the so-called \emph{thin-wall approximation}, which holds 
provided the following two conditions are satisfied:
\begin{enumerate}[(i)]
 \item \label{TW1} The height of the barrier of the potential is much 
larger than the energy difference 
\begin{align}
 \DelU \coloneqq U(\psi_+) - U(\psi_-) 
 \label{epsilon} 
\end{align}
between a false vacuum and another false vacuum, to which the tunnelling occurs.
 \item \label{TW2} The width of the surface wall of the bubble, through which 
the initial false vacuum decays, is much smaller than the bubble size. 
\end{enumerate}
In our case, the condition \eqref{TW1} gives
\begin{equation}
\DelU \ll 2 \lambda^4 \; .
\label{TW1o}
\end{equation}
We examine the remaining condition \eqref{TW2} along the way.

The bounce action for a general potential within the thin-wall approximation has
been presented in \cite{Parke:1982pm}.
In terms of our variables, the bounce action is given in \eqref{BPoa} of 
App.\ \ref{sec:Parke}. 
\begin{subequations}
Defining
\begin{align}
h_0
&\coloneqq
\frac{H_0}{f} \; ,
\qquad
H_0 
\coloneqq
\sqrt{\frac{8\pi G V_1(\phi_\ast \sin \gamma)}{3}}\; ,
\label{H0}
\end{align}
 the bounce action reads as follows:
\begin{equation}
\label{BPo}
  \begin{aligned}
B 
&\simeq
\frac{2 \cdot 27 \pi^2 (8\lambda^2)^4 }{
\sqrt{\left( \DelU - 48 \lambda^4\kappa \right)^2
+ 12 h_0^2 (48 \lambda^4)}
} \\
&\qquad \quad \times
\frac{1}{
\left(
\DelU
+
\sqrt{\left( \DelU - 48 \lambda^4\kappa \right)^2 + 12 h_0^2 (48 \lambda^4)}
\right)^2
-
\left( 48 \lambda^4 \kappa \right)^2
}
\end{aligned}
\end{equation}
\end{subequations}
From \eqref{BPo} we observe that gravitational back-reactions are negligible 
whenever
\begin{equation}
\kappa 
\ll 
\max
\left\{
\frac{\DelU}{48 \lambda^4},
\frac{h_0}{2\lambda^2}
\right\}
\; .
\label{nobg}
\end{equation}
When \eqref{nobg} is satisfied, the bounce action reduces to
\begin{align}
B 
&\simeq
\frac{2 \cdot 27 \pi^2 (8\lambda^2)^4 }{\sqrt{(\DelU)^2
+ 12 h_0^2 (48 \lambda^4)}
\left(
\DelU
+
\sqrt{(\DelU)^2
+ 12 h_0^2 (48 \lambda^4)}
\right)^2
}.
\label{Bnobg}
\end{align}
The demand \eqref{nobg} suggests that expression \eqref{Bnobg} simplifies 
further in two extreme cases: in the 
following subsection we assume that either $\DelU \slash 48 \lambda^4$ is 
much larger than $h_0 \slash 2\lambda^2$ or vice versa. 
\subsection{Flat-space limit}
Let us first look at the situation that space-time can be regarded as flat, 
i.e.\
the effect of the curvature, represented by $h_0$, of the de Sitter space is 
negligible:
\begin{equation}
\frac{\DelU}{48 \lambda^4}
\gg
\frac{h_0}{2\lambda^2} \; ,
\label{flatp}
\end{equation}
which we will refer to as \emph{flat-space limit}. 
In this case, the action \eqref{Bnobg} reduces to the result of 
Coleman \cite{Coleman:1977py}:
\begin{subequations}
\begin{equation}
B
\simeq
B_0
=
\frac{27\pi^2 S_1^4}{2(\DelU)^3} ,
\label{Blimflat}
\end{equation}
where
\begin{equation}
S_1
\coloneqq
2 \int d\zeta
\left(
U_0 (\psi_B)
-
U_0 (\psi_-)
\right) 
= 8 \lambda^2,
\label{S1}
\end{equation}
\end{subequations}
We refer to \eqref{S1a2}  for the explicit calculation in our set-up.
As shown in App.\ \ref{sec:inst}, the thickness of the surface wall
is $\sim 2/\lambda^2$.
Recalling \eqref{TW2}, the thin-wall approximation is valid in the flat-space 
limit if the bubble size $\bar{\rho}$ satisfies
\begin{subequations}
\begin{equation}
\bar{\rho}
=
\bar{\rho}_0
=
\frac{3 S_1}{\DelU}
=
\frac{24\lambda^2}{\DelU}
\gg 2/\lambda^2 \; ,
\label{TWcond}
\end{equation}
which is equivalent to
\begin{equation}
\frac{\DelU}{12 \lambda^4} \ll 1 \;.
\label{TWcond2}
\end{equation}
\end{subequations}
We observe that \eqref{TWcond2} is satisfied due to \eqref{TW1o}.
Note that in the flat-space limit 
the condition \eqref{nobg} for negligible gravitational 
back-reaction reduces to
\begin{equation}
\kappa \ll \frac{\DelU}{48 \lambda^4} \; .
\label{nobgred}
\end{equation}
\subsection{De Sitter limit}
Next, let us look at the opposite limit of the flat-space 
limit \eqref{flatp},
in which the effect of the curvature of the de Sitter space,
represented by $h_0$, is dominant:
\begin{equation}
\frac{\DelU}{48 \lambda^4 h_0} \ll \frac{h_0}{2\lambda^2}\; .
\label{dSlim}
\end{equation}
We refer to this limit as \emph{de Sitter limit}.
In this case the bounce action \eqref{Bnobg} becomes
\begin{equation}
B \simeq \frac{16 \pi^2 \Lambda^2 f}{H_0^3} \; .
\label{nonflatB}
\end{equation}
So far we have kept the inflaton potential general. In order to quantitatively 
discuss constraints arising from a suppressed tunnelling rate, we specify the 
inflaton potential in the next section.
%
%
\section{Examples: Chaotic inflation}
\label{sec:Ex}
%
Let us study examples with an inflaton potential $V_I(\phi)$ given by a 
monomial, i.e.
\begin{equation}
V_I(\phi) = 
V_p(\phi) \coloneqq
\alpha_p \frac{\phi^p}{p!} .
\label{infV}
\end{equation}
In this section, we will work in the unit $M_P\equiv1$. 
Without loss of generality, we take $\alpha_p > 0$ and assume that inflation 
took place when $\phi > 0$.
The associated slow-roll parameters are defined as follows:
\begin{subequations}
\label{eV+etaV}
\begin{align}
\epsilon_V (\phi)
&\coloneqq 
\frac{1}{2} 
\left(
\frac{V_p'}{V_p}
\right)^2
=
\frac{p^2}{2 \phi^2} \;,
\label{eV}\\
\eta_V  (\phi)
&\coloneqq
\frac{V_p''}{V_p}
=
\frac{p(p-1)}{\phi^2} \; .
\label{etaV}
\end{align}
\end{subequations}
In slow-roll inflation, the spectral index $n_s$ and the tensor-to-scalar ratio 
$r$ can be calculated via
\begin{subequations}
\label{ns+r}
\begin{align}
n_s &= 1 - 6 \epsilon_V (\phi_\ast) + 2 \eta_V (\phi_\ast) \; ,
\label{ns}\\
r &= 16 \epsilon_V (\phi_\ast) \; ,
\label{r}
\end{align}
\end{subequations}
where $\ast$ refers to the value when the pivot scale exited the horizon.
The CMB observations constrain $\epsilon_V, |\eta_V| \lesssim \Ord(10^{-2})$ 
through the relations \eqref{ns+r}, see for instance \cite{Ade:2015lrj}.
The number of e-folds $N$ is readily computed to read
\begin{subequations}
\begin{equation}
N(\phi)
=
\left|
\int_{\phi_{end}}^{\phi}
d\phi
\frac{V_p}{V_p'}
\right|
=
\left|
\int_{\phi_{end}}^{\phi}
d\phi
\frac{\phi}{p}
\right|
=
\frac{1}{2p}
\left[
\phi^2
\right]_{\phi_{end}}^{\phi}
=
\frac{1}{2p}
\left(
\phi^2 - \phi_{end}^2
\right) \; ,
\label{N}
\end{equation}
where we define $\phi_{end}$ by the condition
\begin{equation}
\epsilon_V(\phi_{end}) = 1 \; ,
\label{iend}
\end{equation}
which in the examples under consideration gives
\begin{equation}
\phi_{end} = \frac{p}{\sqrt{2}} \;.
\label{phiend}
\end{equation}
\label{Nip}
\end{subequations}
Inserting \eqref{phiend} into \eqref{N}
and solving $\phi_\ast$ 
for a given $N_\ast \coloneqq N(\phi_\ast)$ yields
\begin{equation}
\phi_\ast = 
\sqrt{
2p \left(
N_\ast + \frac{p}{4}
\right)
} \;.
\label{phiast}
\end{equation}
The scalar power spectrum in slow-role inflation is given as
\begin{equation}
P_s
=
\frac{V_p(\phi_\ast)}{24\pi^2 \epsilon_V(\phi_\ast)}
=
2.2 \cdot 10^{-9} \; ,
\label{Ps}
\end{equation}
where the numerical value stems from CMB observations \cite{Ade:2015lrj}.

The coefficient $\alpha_p$ in \eqref{infV}, for a given $N_\ast$, is determined 
by first computing $\phi_\ast$ via \eqref{phiast}, then inserting this value 
into \eqref{Ps} and subsequently solving for $\alpha_p$.
Explicitly,
\begin{subequations}
\label{Pa}
\begin{equation}
P_s 
=
\frac{\alpha_p}{12\pi^2}\frac{\phi_\ast^{p+2}}{p!p^2}
=
2.2 \cdot 10^{-9}  \; ,
\label{Psphi}
\end{equation}
thus
\begin{equation}
\alpha_p 
=
12\pi^2\frac{p!p^2}{\phi_\ast^{p+2}} \cdot P_s  
= 
12\pi^2\frac{p!p^2}{\phi_\ast^{p+2}} \cdot 2.2 \cdot 10^{-9} \; .
\label{alphap}
\end{equation}
\end{subequations}
Now, we use this input data from inflation models constrained by 
CMB observations to estimate the corresponding tunnelling rate in Dante's 
Inferno model.
The parameter $\DelU$, as defined in \eqref{epsilon}, reads in the current 
example as follows:
\begin{align}
\DelU
&=
\frac{1}{f^4}
\bigg(
V_1 (\phi_\ast \sin \gamma)
-
V_1 (-2\pi f \cos \gamma +\phi_\ast \sin \gamma)
\bigg)
\nn\\
&=
\frac{1}{f^4}
\bigg(
V_p (\phi_\ast)
-
V_p (\phi_\ast-2\pi f \cot \gamma)
\bigg)
\nn\\
&\simeq
\cot \gamma
\frac{2\pi V_p'(\phi_\ast)}{f^3} \; ,
\label{eours}
\end{align}
where we have used two ingredients to obtain the last line: Firstly, we 
employed \eqref{fapprox}, more precisely
\begin{equation}
2 \pi f \cot \gamma \simeq 2 \pi f_2 \;  ,
\label{fcot}
\end{equation}
and, secondly, due to the smallness of the slow-roll parameters $\epsilon_V$ 
and 
$\eta_V$, see \eqref{eV+etaV}, it follows that the inflaton potential 
$V_p(\phi)$ around $\phi \sim \phi_\ast$ does not change much over the Planck 
scale, i.e.\ $M_P \gtrsim 2\pi f_2$.

In the following two subsections we examine the tunnelling rate
in two scenarios: Firstly, in the flat-space limit and, secondly, in the de 
Sitter limit.
%
%
\subsection{Flat-space limit}
\label{subsec:flat}
%
We begin with the parameter region of Dante's Inferno model in which the 
flat-space limit \eqref{flatp} is appropriate, i.e.\
\begin{equation}
\frac{\DelU}{48 \lambda^4}
\gg
\frac{h_0}{2\lambda^2} \; .
\label{flatp2}
\end{equation}
For negligible gravitational back-reaction the bounce action in the flat-space 
limit is provided in \eqref{Blimflat}.
We investigate the validity of the negligibility of the gravitational 
back-reaction later in the subsection.
Inserting \eqref{eours} into \eqref{Blimflat} yields
\begin{equation}
B 
= 
\frac{27 \cdot 2^8 \Lambda^8 f}{\pi}
\left(
\frac{\tan \gamma}{V_p'(\phi_\ast)}
\right)^3 \; ,
\label{pB}
\end{equation}
where we have used $S_1 = 8 \lambda^2$ for our set-up (c.f.\ \eqref{S1a2} in 
App.\ \ref{sec:inst}).
Then, for a suppressed tunnelling process, i.e.\ $B \gg 1$, the following 
condition has to hold:
\begin{subequations}
\begin{equation}
\tan \gamma 
\gg
\left(
\frac{27 \cdot 2^8 \Lambda^8 f}{\pi}
\right)^{-1/3}
V_p'(\phi_\ast)
\eqqcolon
\tan \gamma_{T}.
\label{sT}
\end{equation}
Using \eqref{Nip}--\eqref{Pa}, the explicit form of $\tan \gamma_T$ reads as
\begin{equation}
\tan \gamma_T 
= 
2^{5/6} \pi^{7/3}   \cdot \frac{1}{\left(f \Lambda^8\right)^{1/3}} 
\cdot \left[ P_s
\left(\frac{p}{4 N_\ast+p}\right)^{3/2} \right] \;.
\label{tTpN}
\end{equation}
\end{subequations}
In \eqref{tTpN} the numerical factor in the squared brackets is determined by 
the parameters of the inflation model $p,N_\ast$, and CMB observations 
\eqref{Ps}.
The constraint \eqref{sT} should be compared with the defining condition of 
Dante's 
Inferno model \eqref{slopes}, which in terms of the parameters of the model 
gives
\begin{subequations}
\begin{equation}
\frac{\Lambda^4}{f}
\gg
\cot \gamma
V_p'(\phi_\ast)\; ,
\label{sDIpre}
\end{equation}
or equivalently
\begin{equation}
\tan \gamma 
\gg 
\frac{f}{\Lambda^4}
V_p'(\phi_\ast)
\eqqcolon
\tan \gamma_{DI}  \; .
\label{sDI}
\end{equation}
\end{subequations}
In the above, we have used
\begin{equation}
\frac{dV_p}{d\phi}(\phi)
=
\frac{d}{d\phi} V_{1} (\phi \sin \gamma)
=
\frac{d\phi_1}{d\phi}
\frac{dV_1}{d\phi_1} (\phi_1 = \phi \sin \gamma)
=
\sin \gamma
\frac{dV_1}{d\phi_1}(\phi_1) \; ,
\label{pdV}
\end{equation}
which follows from \eqref{VI} and the usual chain rule.
Again, using \eqref{Nip}--\eqref{Pa} allows to specialise $\tan \gamma_{DI}$ to
\begin{equation}
\tan \gamma_{DI}
=
24 \sqrt{2} \pi^2   \cdot 
\frac{f}{\Lambda^4}
\cdot
\left[ P_s
\left(\frac{p}{4 N_\ast +p}\right)^{3/2}
\right] \; .
\label{tDIpN}
\end{equation}
We are particularly interested in the scenario for which condition \eqref{sT} 
for suppressed tunnelling enforces a stronger condition on the model 
than \eqref{DI2}.
From \eqref{sT} and \eqref{sDI}, this is the case for
\begin{equation}
\tan \gamma_{T} > \tan \gamma_{DI} \; .
\label{T}
\end{equation}
In terms of the parameters of the model under consideration, the 
inequality \eqref{T} reduces to
\begin{subequations}
	\begin{equation}
	\left(
	\frac{27 \cdot 2^8 \Lambda^8 f}{\pi}
	\right)^{-1/3} 
	>
	\frac{f}{\Lambda^4} \; ,
	\label{snoT2pre}
	\end{equation}
	or equivalently
	\begin{equation}
	\frac{\Lambda}{f}
	\gtrsim 7 \;  .
	\label{snoT}
	\end{equation}%
\end{subequations}
Whenever \eqref{snoT} is satisfied,
the constraint \eqref{sT}, which ensures the suppression of the tunnelling, is 
more restrictive than the defining condition \eqref{slopes} of Dante's Inferno 
model. In other words, in the region of the parameter space where \eqref{snoT} 
holds tunnelling is not automatically suppressed in Dante's Inferno model; 
thus, an additional constraint arises\footnote{As discussed in the beginning 
of Sec.\ \ref{sec:TDI}, we only 
consider the region $f_1, f_2, \Lambda \gtrsim H$.}. 

Note that in Dante's Inferno model derived from a higher dimensional gauge 
theory discussed in Sec.\ \ref{subsec:HDG}, \eqref{snoT} assures that
the tunnelling process is suppressed for natural values of the 
parameters \eqref{HDGLf} (at least in the simplest version of the model).

To demonstrate how condition \eqref{sT} constrains the parameter space,
we illustrate the scenarios $\Lambda/f = 10$, $p=1,2$, and $N_\ast = 60$ in
Fig.\ \ref{fig:10fp1} and Fig.\ \ref{fig:10fp2}, respectively.
If one wishes to fix a certain value of $\tan \gamma$ then the
condition \eqref{sT} yields a lower bound on $f \simeq f_1$.
For example, if we demand $\tan \gamma \sim 5 \cdot 10^{-2}$ then $f \gtrsim 
10^{-4}$ is required in the above cases,
as can be read off from \eqref{tTpN} or Fig.\ \ref{fig:10fp1} and 
Fig.\ \ref{fig:10fp2}.

Now, let us focus on condition \eqref{flatp2}, which defines 
the flat-space limit.
From \eqref{le}, \eqref{H0}, and \eqref{eours}, one infers that 
condition \eqref{flatp2} becomes
\begin{subequations}
\begin{equation}
\frac{\DelU}{24 \lambda^2 h_0}
=
\cot \gamma
\frac{\pi}{12}
\frac{V_p'(\phi_\ast)}{\Lambda^2 H_0}
\gg 1 \; ,
\label{condflat}
\end{equation}
which we recast as
\begin{equation}
\tan \gamma 
\ll
F 
\coloneqq
\frac{\pi}{12} \frac{V_p'(\phi_\ast)}{\Lambda^2 H_0}
\; .
\label{singF2}
\end{equation}
Specialising $F$ via \eqref{Nip}--\eqref{Pa} to the current model, we obtain
\begin{equation}
F
=
\frac{1}{\Lambda^2}
\cdot
\left(\pi^2  \sqrt{P_s} \cdot
\frac{ p }{4 N_\ast+p}
\right)
\sim
\frac{1}{\Lambda^2}
\cdot
\Ord(10^{-6}),
\label{FpN}
\end{equation}
\end{subequations}
where the last numerical value holds for $p=1,2$ with $N_\ast = 50-60$.
As discussed around \eqref{sing}, Dante's Inferno model requires
$\tan \gamma \lesssim \Ord(10^{-1})$ or less.
Thus, the flat-space limit is appropriate for 
\begin{equation}
\Lambda \ll \Ord(10^{-5}) \; .
\label{Flim}
\end{equation}
From \eqref{singF2} and \eqref{sDI}, one readily computes the following ratio:
\begin{equation}
\frac{F}{\tan \gamma_{DI}}
=
\frac{\pi}{12} \frac{\Lambda^2}{f H_0}
=
 \frac{1}{24 \sqrt{2 P_s}} \sqrt{\frac{4N_\ast +p}{p }} \cdot \frac{\Lambda^2}{ 
f}
\sim 
\Ord(10^5) \cdot 
\left(\frac{\Lambda}{f} \right) \Lambda \; ,
\label{FsDI}
\end{equation}
where we have used $p=1,2$ and $N_\ast \sim 50-60$.
Hence, when $\Lambda/f \gtrsim 7$ as in \eqref{snoT}, then \eqref{FsDI} implies 
$F \gg \tan \gamma_{DI}$, provided $\Lambda \gtrsim \Ord(10^{-5})$ holds.

Next, we examine condition \eqref{nobgred} for negligible gravitational 
back-reaction. 
In terms of the parameters of the current model, we obtain
\begin{subequations}
\begin{equation}
\tan \gamma 
\ll 
K\coloneqq
\frac{\pi}{24}
\frac{V_p'(\phi_\ast)}{\Lambda^2 f^3} \; .
\label{K}
\end{equation}
By means of \eqref{Nip}--\eqref{Pa}, we explicitly parametrise $K$ as
\begin{equation}
K
= 
\frac{1}{\Lambda^2 f^3}
\cdot
\left[
{\sqrt{2} \pi^3 {P_s}} 
\cdot  
\left(
 \frac{p}{4N_\ast+p}
\right)^{3/2}
\right]
\sim
\frac{1}{\Lambda^2 f^3}
\cdot
\Ord(10^{-11}) \; ,
\label{KpN}
\end{equation}
\end{subequations}
where the last numerical value holds for $p=1,2$ with $N_\ast = 50-60$.
As we assume $\tan \gamma \ll 1$ in Dante's Inferno model, if $K \gtrsim 1$ 
then \eqref{K} does not introduce a further constraint.
Thus, $K \gtrsim 1$ whenever
\begin{equation}
\Lambda^2 f^3 \lesssim \Ord(10^{-11}) \;.
\end{equation}
For the range of the parameter $f$ as displayed in Fig.\ \ref{fig:La10p1} to 
Fig.\ \ref{fig:La10p2}, $K$ is always much greater than $1$ and, therefore,
the gravitational back-reaction can be neglected.

We note that \eqref{K} and \eqref{singF2} imply the following ratio:
\begin{equation}
\frac{K}{F}
=
\frac{H_0}{2f^3}.
\label{KF}
\end{equation}
Then $H_0 \sim \Ord(10^{-5})$, for $p=1,2$ with $N_\ast = 50-60$ as previously 
used, implies that $K \gtrsim F$ for $f \lesssim \Ord(10^{-2})$.
In this case, $\tan \gamma \ll K$ is automatically satisfied if $\tan \gamma 
\ll 
F$.
%
%

Finally, we verify the validity of the thin-wall approximation.
Inserting \eqref{eours} and \eqref{le} into the condition \eqref{TWcond} 
for the validity of the thin-wall approximation gives
\begin{equation}
\frac{6}{\pi} \frac{\Lambda^4}{f V_1'} \gg 1 .
\label{twours}
\end{equation}
We have used \eqref{pdV} to obtain \eqref{twours}.
Using the parameter $s$, as introduced in \eqref{slopes}, one can 
rewrite \eqref{twours} as
\begin{equation}
\frac{6}{\pi} s
\gg
1 \; .
\label{twours2}
\end{equation}
We recall that $s\gg 1$ is one of the conditions \eqref{slopes} required in 
Dante's Inferno model.
Thus, the condition \eqref{twours2} for the validity of the thin-wall 
approximation gives numerically the same constraint on Dante's Inferno model 
as \eqref{slopes}, up to a minor difference of a $\Ord(1)$ numerical factor.
As a consequence, the thin-wall approximation is always valid in Dante's 
Inferno model in the flat-space limit.%
%
\begin{figure}[h]
\centering 
\begin{subfigure}[b]{0.49\textwidth}
\begin{tikzpicture}
\begin{semilogxaxis}[
 width=1\textwidth,
 xlabel = $f$,
 xmin = 1e-5, xmax=1e-3,
 ymin = 0, ymax = 0.10,
 restrict y to domain = 0:0.15
]
\addplot [
domain = 1e-5:1e-3,
samples = 50,
color = red
]
{ (1.9698*10^(-10)* x)/(10*x)^4
};
\addlegendentry{$\tan( \gamma_{DI})$}
\addplot [
domain = 1e-5:1e-3,
samples = 50,
color = blue
]
{
1.51451*10^(-11)/(x* (10*x)^8)^(1/3)
};
\addlegendentry{$\tan (\gamma_{T})$}
\end{semilogxaxis}
\end{tikzpicture}
\caption{ }
\label{fig:10fp1}
\end{subfigure}
\begin{subfigure}[b]{0.49\textwidth}
\begin{tikzpicture}
\begin{semilogxaxis}[
 width=1\textwidth,
 xlabel = $f$,
 xmin = 1e-5, xmax=1e-3,
 ymin = 0,   ymax = 0.30,
 restrict y to domain = -0.2:0.35,
 legend pos=north west
]
\addplot [
domain = 1e-5:1e-3,
samples = 50,
color = black
]
{ 1.92085*10^(-6)/(10*x)^2
};
\addlegendentry{$F$}
\end{semilogxaxis}
\end{tikzpicture}
\caption{ }
\label{fig:F10fp1}
\end{subfigure}
\caption{The exemplary parameter values are $p=1$, $N=60$, and $\Lambda=10f$.}
\label{fig:La10p1}
\end{figure}
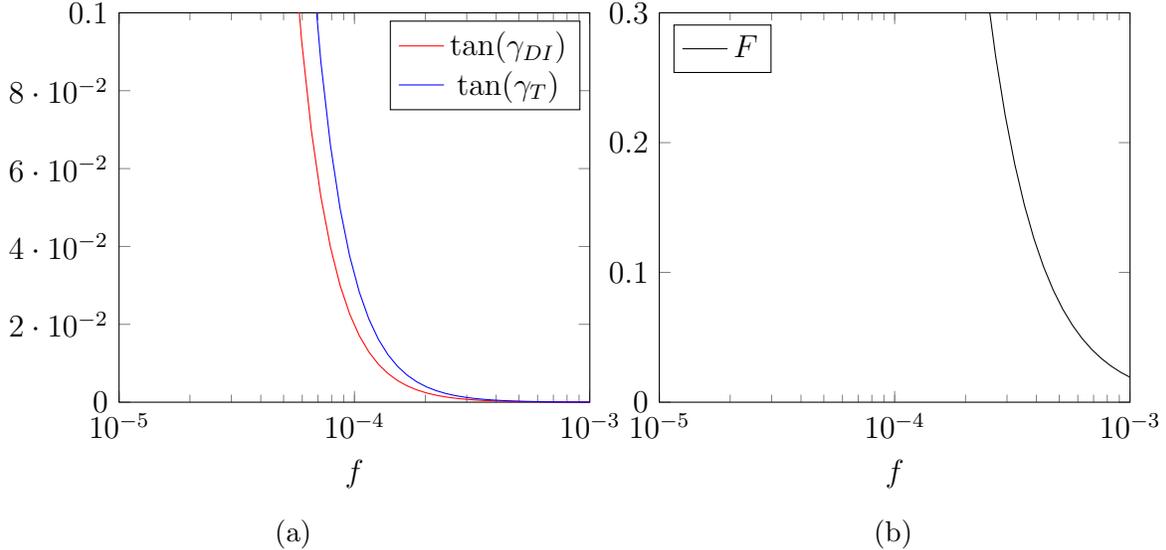
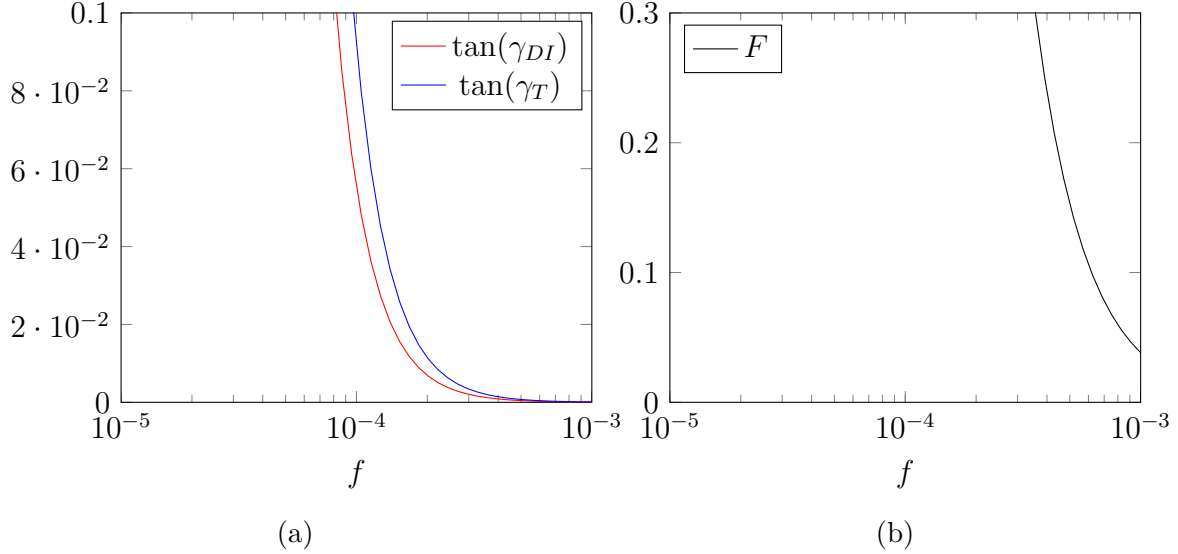
\begin{figure}[h]
\centering
\begin{subfigure}[b]{0.49\textwidth}
\begin{tikzpicture}
\begin{semilogxaxis}[
width=1\textwidth,
 xlabel = $f$,
 xmin = 1e-5, xmax=1e-3,
 ymin = 0 ,  ymax = 0.1 ,
 restrict y to domain = -0.2:0.15
]
\addplot [
domain = 1e-5:1e-3,
samples = 50,
color = red
]
{ (5.53695*10^(-10)*x)/(10*x)^4 
};
\addlegendentry{$\tan( \gamma_{DI})$}
\addplot [
domain = 1e-5:1e-3,
samples = 50,
color = blue
]
{4.25715*10^(-11)/(x* (10*x)^8)^(1/3)
};
\addlegendentry{$\tan (\gamma_{T})$}
\end{semilogxaxis}
\end{tikzpicture}
\caption{ }
\label{fig:10fp2}
\end{subfigure}
\begin{subfigure}[b]{0.49\textwidth}
\centering
\begin{tikzpicture}
\begin{semilogxaxis}[
width=1\textwidth,
 xlabel = $f$,
 xmin = 1e-5, xmax=1e-3,
 ymin = 0,  ymax = 0.30,
 restrict y to domain = -0.2:0.35,
 legend pos=north west
]
\addplot [
domain = 1e-5:1e-3,
samples = 50,
color = black
]
{ 3.82583*10^(-6)/(10*x)^2
};
\addlegendentry{$F$}
\end{semilogxaxis}
\end{tikzpicture}
\caption{ }
\label{fig:F10fp2}
\end{subfigure}
\caption{The exemplary parameter values are $p=2$, $N=60$, and $\Lambda=10f$.}
\label{fig:La10p2}
\end{figure}
Finally, we notice from \eqref{pB} that, within the class of monomial inflation 
potentials, the tunnelling rate either stays constant (for $p=1$) or decreases 
(for $p>1$) as $\phi_\ast$ decreases.
Therefore, it is sufficient to estimating the tunnelling rate at 
$\phi=\phi_\ast$ in order to verify the suppression of the tunnelling process 
in these cases.
%
%
\subsection{De Sitter limit}
\label{subsec:dS}
%
In this subsection, we study the de Sitter limit \eqref{dSlim} which gives
\begin{subequations}
\begin{equation}
\frac{\DelU}{24 \lambda^2 h_0}
=
\cot \gamma
\frac{\pi}{12}
\frac{V_p'(\phi_\ast)}{\Lambda^2 H_0}
\ll 1 \; ,
\label{conddS}
\end{equation}
or equivalently
\begin{equation}
\tan \gamma \gg 
F:=
\frac{\pi}{12}
\frac{V_p'(\phi_\ast)}{\Lambda^2 H_0}
\sim \frac{1}{\Lambda^2} \cdot \Ord(10^{-6}) \; ,
\label{FdSlim}
\end{equation}
\end{subequations}
where the last approximation holds for $p=1,2$ with $N_\ast \sim 50-60$.
As discussed around \eqref{sing}, Dante's Inferno model requires $\tan \gamma 
\lesssim \Ord(10^{-1})$ or less.
Then \eqref{FdSlim} implies at least
\begin{equation}
\Lambda^2 \gg \Ord(10^{-5}) \; .
\label{dSLambda}
\end{equation}
One should keep in mind that the right hand side of \eqref{dSLambda} can be 
even smaller, depending on the desired $\tan \gamma$.

In the de Sitter limit \eqref{conddS}, the bounce is given by \eqref{nonflatB}
when gravitational back-reaction is negligible.
We will examine gravitational back-reaction shortly.
In this case, the condition for a suppressed tunnelling rate, i.e.\ $B \gg 1$, 
becomes
\begin{equation}
\Lambda^2 f \gg \frac{H_0^3}{16\pi^2} 
\sim \Ord(10^{-16}) \; ,
\label{nonflatTcond}
\end{equation}
where we have used the value of $H_0$ for $p=1,2$ with $N_\ast = 50-60$.
In the parameter region of a sufficiently rapid pre-factor $A$, i.e.\ all 
relevant scales are above the Hubble scale at the time of inflation, condition 
\eqref{nonflatTcond} is not a constraint at all. Therefore, the tunnelling 
process is suppressed provided the gravitational back-reaction is negligible 
and the thin-wall approximation is applicable. 

Consequently, we focus on the gravitational back-reaction for the de Sitter 
limit first.
In this case, condition \eqref{nobg} for negligibility of the gravitational 
back-reaction reads
\begin{subequations}
\begin{equation}
\kappa \ll \frac{h_0}{2\lambda^2} \; .
\label{dSnobg}
\end{equation}
In terms of the original parameters of the model, \eqref{dSnobg}
becomes
\begin{equation}
2 \Lambda^2 f \ll H_0 \sim \Ord(10^{-5}).
\label{dSnobgo}
\end{equation}
By means of \eqref{dSLambda}, condition \eqref{dSnobgo} implies
\begin{equation}
f \ll \Ord(1) \; .
\label{dSf}
\end{equation}
\end{subequations}
Reminding ourselves of one of the defining conditions of Dante's Inferno 
model \eqref{DI1}, we conclude that \eqref{dSf} is always satisfied
in this model.
Therefore, in Dante's Inferno model, the gravitational back-reaction is always 
negligible in the de Sitter limit.

Finally, let us verify the validity of the thin-wall approximation in the de 
Sitter limit.
To begin with, we note that the bubble size $\bar{\rho}$ in the 
de Sitter limit is always smaller than the bubble size $\bar{\rho}_0$ in
the flat-space limit, which follows from the definition \eqref{barrho} of 
$\bar{\rho}$ in App.\ \ref{sec:Parke} and the fact that
$x$ and $y$ in \eqref{xy} are positive numbers.

For a quantitative estimate of $\bar{\rho}$, we specialise
$x$ and $y$ of \eqref{xy} to the parameters in our model
(see \eqref{xyo}):
\begin{subequations}
\begin{align}
x &=
\kappa \cdot
\left(
\frac{\DelU}{48 \lambda^4}
\right)^{-1} \; ,
\label{x}  \\
y 
&=
\frac{6 h_0^2}{\kappa \DelU} + 1
\simeq \frac{6 h_0^2}{\kappa \DelU} \; .
\label{y}
\end{align}
\label{xyours}
\end{subequations}
The last approximation in \eqref{y} always holds in the de Sitter limit.
From \eqref{xyours} we immediately infer
\begin{subequations}
\begin{equation}
2xy 
\simeq
\left(
\frac{\DelU}{24\lambda^2h_0}
\right)^{-2} \gg 1 \; ,
\label{xyp}
\end{equation}
where the last hierarchy is a consequence of the de Sitter limit \eqref{dSlim}.
Moreover, the negligible gravitational back-reaction in the de Sitter limit, as 
discussed in \eqref{dSf}, allows to deduce  
\begin{equation}
\frac{x}{y}
=
2\kappa^2 \cdot
\left(
\frac{h_0}{2\lambda^2}
\right)^{-2} \ll 2 \; .
\label{xoy}
\end{equation}
\end{subequations}
By means of \eqref{xyp} and \eqref{xoy}, we then obtain
\begin{equation}
\bar{\rho}^2
\simeq
\frac{\bar{\rho}_0^2}{2xy}
=
\left(
\frac{24 \lambda^2}{\DelU}
\right)^2
\left(
\frac{\DelU}{24\lambda^2h_0}
\right)^{2}
=
\frac{1}{h_0^2}.
\label{dSbarrho}
\end{equation}
Since the thickness of the surface wall is given as $\sim 2/\lambda^2$ as 
described in App.\ \ref{sec:inst}, the second condition \eqref{TW2} of the 
thin-wall approximation becomes 
\begin{equation}
\frac{\lambda^2 \bar{\rho}}{2}
\simeq
\frac{\lambda^2}{2h_0} 
=
\frac{\Lambda^2}{2H_0 f}
\sim
\frac{\Lambda^2}{2f}
\cdot
\Ord(10^5)
\gg 1,
\label{dSTW}
\end{equation}
where we have used the value of $H_0$ for $p=1,2$ and $N_\ast = 50-60$.
Consequently, \eqref{dSLambda} and \eqref{DI1} imply that \eqref{dSTW} 
is always satisfied in the current model within the de Sitter limit. 
Therefore, the thin-wall approximation is always appropriate in this limit.
%
\section{Summary and discussions}\label{sec:discussions}
%
In this article, we studied tunnelling in Dante's Inferno model
within the thin-wall approximation and subsequent constraints on the parameter 
space.

In general, we argued that the tunnelling process can only become fatal for 
inflation if all scales in Dante's model satisfy $\Lambda,f_1,f_2\gtrsim H$, 
and 
if $B$ is less than order one. All other parameter regions are intrinsically 
safe from tunnelling in the leading order of $\hbar$.

We have shown in \eqref{Flim} that the flat-space limit is appropriate for 
$\Lambda \ll \Ord (10^{-5})$.
In the flat-space limit, the parameter space is simultaneously constrained by 
the condition \eqref{sT} for a suppressed tunnelling rate, and one of the 
defining conditions \eqref{sDI} of Dante's Inferno model, i.e.
\begin{equation}
\tan \gamma = \frac{f_1}{f_2} 
\gg \max\{ \tan \gamma_T, \tan \gamma_{DI} \} \; .
\label{tanallowed}
\end{equation}
We have seen that for a fixed ratio $\Lambda/f$, a lower bound on the parameter 
$f \simeq f_1$ is imposed by \eqref{tanallowed}, for a given $f_1/f_2$.
In particular, we have shown in \eqref{snoT} that 
$\tan \gamma_T$ is bigger than $\tan \gamma_{DI}$
when $\Lambda/f \gtrsim 7$,
in which case the condition for a suppressed tunnelling rate
gives a stronger constraint than the defining condition
of Dante's Inferno model.
Since the parameter space is multi-dimensional, one has to choose certain 
parameters to obtain a visualisable subspace. 
We computed the bounds numerically in monomial chaotic inflation with
$\Lambda/f = 10$,
$p=1,2$ with $N_\ast = 50-60$ and exemplified these in Fig.\ 
\ref{fig:La10p1} -- \ref{fig:La10p2}. 
While numerical values of the bounds were given in the examples, the method for 
obtaining the bound is clearly general and can be straightforwardly applied to
other forms of the inflaton potential.

In the de Sitter limit
which was shown to be appropriate for $\Lambda^2 \gg \Ord(10^{-5})$
in \eqref{dSLambda},
the condition for a suppressed tunnelling rate is 
trivial in the problematic region $\Lambda,f_1,f_2,\gtrsim H$.
In other words, the tunnelling process is
always suppressed in this limit.

We summarized those constraints on the parameter space in
Table.\ \ref{tab:numerical_bounds}.

Additionally, we identified in each limit the parameter region in which the 
thin-wall approximation is valid and the gravitational back-reactions are 
negligible. 
It turned out that this covers a large part of the parameter region of interest.
\begin{table}
\centering
\begin{tabular}{|c|c|c|}
\hline
& flat-space limit & de Sitter limit \\ 
\hline
valid for & $\Lambda^2 \ll \Ord(10^{-5})$ &  $\Lambda^2 \gg
\Ord(10^{-5})$ \\
constraints & $\tan \gamma \gg \tan \gamma_T$
when ${\Lambda}/{f}\gtrsim 7$  & no constraint \\ 
\hline
\end{tabular}
\caption{A brief summary of constraints on the parameter space.}
\label{tab:numerical_bounds}
\end{table}

The original article \cite{Berg:2009tg} mentioned that a useful value of
$\Lambda$ lie in the range from $10^{-3} M_P$ to $10^{-1} M_P$, 
and typical 
values for $f_1$ and $f_2$ are $10^{-3} M_P$ and $10^{-1} M_P$, respectively.
Our results confirm that these values are safe from tunnelling,
and further provide explicit constrains on these parameters from the condition
of a suppressed tunnelling rate.

For the article at hand, we restricted ourselves to the level of an effective 
field theory.
For example, the parameter $\Lambda$ in \eqref{VDI}, which controls the height 
of the sinusoidal potential, was treated as an input parameter.
However, when the model is embedded in a UV theory
the height of the sinusoidal potential could be a function of the
required monodromy number $N_{mon} \coloneqq \Delta \phi  \cdot \cos \gamma 
\slash (2\pi f_2) \sim \Ord(10) \cdot M_P / f_2$. 
(Here $\Delta \phi$ denotes the field distance the inflaton field travels 
during the inflation.)
In a related model embedded in string theory, for instance, the height of the 
sinusoidal potential was shown \cite{McAllister:2016vzi} to be proportional to
$ e^{-\gamma_{br} N_{mon}}$, where $\gamma_{br}$ is a parameter independent of 
$N_{mon}$.
Such a rapid decrease of the height of the sinusoidal potential for increasing 
$N_{mon}$ would give rise to much severer constraint on the axion decay 
constants than the ones given in this article.
Consequently, UV completions of Dante's Inferno model and the constraints from 
it are certainly an important direction to be investigated in the future.
\vspace*{8mm}
\begin{center}
\noindent{\bf Acknowledgments}
\end{center}
We would like to thank Yoji Koyama, Olaf Lechtenfeld, and 
Marco Zagermann for useful discussions.
This collaboration was supported by a Short Term Scientific Mission (STSM)
under COST action MP1405. 
MS was supported by the DFG research training group GRK1463 ``Analysis, 
Geometry, and String Theory'' and the Insitut für Theoretische Physik of the 
Leibniz Universität Hannover. MS is currently supported by Austrian Science 
Fund (FWF) grant P28590. 
MS would like to thank Manipal Center for Natural Sciences,
Manipal University for hospitality and support during the visit.\\[6mm]
%
%
\appendix
\section{Effects of time evolution of inflaton on tunnelling}
\label{sec:constant_field}
In this appendix we justify the claim of regarding $\phi$ as being fixed in 
time during inflation. 
Accounting for changes of the tunneling rate through a time variation of 
$\phi(\xi)$ is achieved by modifying \eqref{SE_explicit} as follows:
\begin{align}
U_1(\psi,\zeta)
&\coloneqq
\frac{1}{f^4} V_1 (f \psi \cos \gamma + \phi(\zeta) \sin \gamma) \; 
\label{U1t} \; ,
\end{align}
i.e.\ one simply keeps the time dependent $\phi(\zeta)$ instead of choosing the 
reference point $\phi_\ast$.
\subsection{Flat-space limit}
In the flat-space limit
discussed in Sec.\ \ref{subsec:flat}, 
the time variation of $U_1$ may become relevant 
through its appearance in $\Delta U$. 
As in \eqref{eours}, $\Delta U$ can be estimated as
\begin{align}
\DelU
\coloneqq
U(\psi_+) - U(\psi_-)
\simeq
\cot \gamma
\frac{2\pi V_I'(\phi(\xi))}{f^3} 
\label{eourst} \; ,
\end{align}
where $V_I$ is defined in \eqref{VI}.

To judge the impact of a time dependent $\phi$, we examine the time variation 
of $\Delta U$ in a time interval $\Delta \zeta$ relative to $\Delta U$ itself. 
In detail
\begin{align}
\frac{1}{\Delta U} \cdot \frac{\diff \Delta U}{\diff \zeta} 
\cdot 
\Delta \zeta
\simeq
 \frac{V_I''}{V_I'} \cdot  \frac{\diff}{\diff \xi} \phi(\xi) \cdot \Delta \xi
\simeq
\eta_V H \Delta \xi \; ,
\label{tchange}
\end{align}
where we used the slow-role approximation of the equations of motion, i.e.\
\begin{align}
3 H \frac{\diff \phi}{\diff \xi} 
\simeq 
V_I' 
\; , \qquad 
3 H^2 
\simeq V_I \; .
 \label{eq:eom_slow_role}
\end{align}
Recall Sec.\ \ref{sec:TDI}, we assume that  
\emph{all} relevant physical parameters are greater than the
Hubble expansion rate $H$, i.e.\ 
$\Lambda, f_1, f_2 \gtrsim H$. 
Consequently, during a time interval $\Delta \xi \simeq 1/H$ the relative 
change \eqref{tchange} becomes of order $\eta_V$, and we may safely neglect the 
time dependence of $\DelU$ in the slow-role regime $\eta_V \lesssim 
\Ord(10^{-2})$.
\subsection{de Sitter limit}
In de Sitter limit
discussed in 
Sec.\ \ref{subsec:dS},
a time variation of $U_1$ enters via the time variation of 
$V_I$.
In slow-role inflation models, it is well-known that the time variation of 
$V_I$ is suppressed by the slow-role parameters. To be explicit, we find
\begin{align}
\frac{1}{V_I}
\cdot
\frac{\diff}{\diff \zeta} V_I 
\cdot \Delta \zeta
\simeq
\frac{V_I'}{V_I}\cdot  \frac{\diff \phi}{\diff \zeta} \cdot \Delta \zeta 
\simeq
2 \epsilon_V H \Delta \xi \; ,
\label{VIchange}
\end{align}
where we again made use of \eqref{eq:eom_slow_role}.
Hence, \eqref{VIchange} is small in a time-scale $\Delta \xi \sim 1/H$,
as $\epsilon_V \lesssim \Ord(10^{-2})$.
%
\section{Summary of the bounce solution in the thin-wall approximation}
\label{sec:Parke}
%
The bounce action for general potential in the thin-wall approximation has been 
given in \cite{Parke:1982pm} (\cite{Sasaki:1994yj} is also a useful read).
Here, we review the necessary results.
The relevant Euclidean action is of the form
\begin{equation}
S_E
=
2\pi^2
\int
d\zeta
\rho^3
\left(
\frac{1}{2}
\dot{\psi}^2
+
U(\psi)
\right)
-
\frac{3\rho}{\kappa_P}
\left(
\dot{\rho}^2+1
\right).
\label{SEP}
\end{equation}
The action \eqref{SEP} has the same form\footnote{%
Precisely speaking, variables in \eqref{SErho} were dimensionless,
but it is straightforward to implement this point in the comparison,
e.g. by setting $M_P \equiv 1$ as we did.} as \eqref{SErho},
with parameters $\kappa$ being replaced with $\kappa_P$ defined by
\begin{equation}
\kappa_P \coloneqq 8\pi G = M_P^{-2} \;.
\label{kPH0}
\end{equation}
The bounce action is introduced as 
\begin{equation}
B\coloneqq S_E[\psi_B] - S_E[\psi_+] \; ,
\label{Ba}
\end{equation}
where $\psi_B$ is the bounce solution.
In the thin-wall approximation, which is appropriate whenever 
conditions \eqref{TW1} and \eqref{TW2} hold, we evaluate \eqref{Ba} by dividing 
the integration region into three parts:
Outside the bubble, at the surface of the bubble, and inside the bubble.
Outside the bubble, the bounce and false vacuum are identical;
therefore, the contribution $B_{out}$ to the bounce action is
\begin{equation}
B_{out} = 0 \;.
\label{Bout}
\end{equation}
At the surface wall of the bubble,
we can replace $\rho$ by the position of the 
centre of the surface wall
$\bar{\rho}$. 
Then, the contribution to the bounce action from 
the surface wall $B_w$ is given by
\begin{subequations}
\begin{equation}
B_{w}
=
2 \pi^2 \bar{\rho}^3 S_1 \; ,
\label{Bw}
\end{equation}
where
\begin{equation}
S_1
\coloneqq
2 \int d\zeta
\left(
U_0 (\psi_B)
-
U_0 (\psi_-)
\right) \; .
\label{S1a}
\end{equation}
\end{subequations}
Inside the bubble, $\psi$ is constant, such that \eqref{rhoeq} allows to 
deduce 
\begin{subequations}
\begin{equation}
d\zeta 
=
d \rho
\left(
1 - \frac{\kappa_P}{3}\rho^2 U(\psi)
\right)^{1/2}  \; ,
\label{measure}
\end{equation}
and we then obtain
\begin{align}
B_{in}
&=
-\frac{12\pi^2}{\kappa_P}
\int_0^{\bar{\rho}} d\rho
\left[
\left(
1-\frac{\kappa_P}{3}U_-
\right)^{1/2}
-
\left(
1-\frac{\kappa_P}{3}U_+
\right)^{1/2}
\right]
\nn\\
&=
\frac{12\pi^2}{\kappa_P^2}
\left[
U_-^{-1}
\left(
\left(1-\frac{\kappa_P \rho^2}{3} U_-\right)^{3/2}
-1
\right)
-
U_+^{-1}
\left(
\left(1-\frac{\kappa_P \rho^2}{3} U_+\right)^{3/2}
-1
\right)
\right] \, ,
\label{Bin}
\end{align}
\end{subequations}
where $U_+\coloneqq U(\psi_+)$ and $U_- \coloneqq U(\psi_-)$ are the energy 
density of the false vacuum we start with and another false vacuum we end with, 
respectively.

Extremising $B$ with respect to $\bar{\rho}$ gives
\begin{subequations}
\begin{equation}
\bar{\rho}^2
=
\frac{
\bar{\rho}_0^2
}
{
1+ 2xy +x^2
} \; ,
\label{barrho}
\end{equation}
where
\begin{equation}
\bar{\rho}_0
\coloneqq
\frac{3 S_1}{(U_+ - U_-)} ,
\label{rho0}
\end{equation}
is the critical bubble size without the presence of gravity, and $x$ and $y$ 
are 
defined as follows:
\begin{equation}
x \coloneqq
\frac{\bar{\rho}_0^2}{4}
\frac{\kappa_P (U_+ - U_-)}{3},
\quad
y \coloneqq
\frac{U_+ + U_-}{U_+ - U_-} .
\label{xy}
\end{equation}
\end{subequations}
The bounce action is obtained as
\begin{subequations}
\begin{equation}
B \simeq
B_0 r(x,y) \; ,
\label{BP}
\end{equation}
where
\begin{equation}
B_0 \coloneqq
\frac{27 \pi^2 S_1^4}{2 (U_+-U_-)^3} \; ,
\label{B0P}
\end{equation}
which is the bounce action in flat space. The function $r(x,y)$ is defined 
as follows:
\begin{align}
r(x,y)
&\coloneqq
2 \cdot \frac{  (1+xy)-\sqrt{1+2xy+x^2}}{x^2  (y^2-1) \sqrt{1+2xy+x^2}}
\nn\\
&=
\frac{2 \cdot \left( (1+xy)^2- (1+2xy+x^2)\right)}{x^2  (y^2-1) 
\sqrt{1+2xy+x^2} \left((1+xy)+\sqrt{1+2xy+x^2}\right)}
\nn\\
&=
\frac{2}{\sqrt{1+2xy+x^2} \left((1+xy)+\sqrt{1+2xy+x^2}\right)} \; .
\label{rP}
\end{align}
\end{subequations}
%
To obtain the bounce action for set-up of this article, one simply has to
replace $\kappa_P$ with $\kappa$ as mentioned earlier.
Then, we use the result \eqref{S1a2} of App.\ \ref{sec:inst}:
\begin{equation}
S_1 = 8 \lambda^2 \; ,
\label{S1o}
\end{equation}
together with
\begin{equation}
U_+ - U_- = \DelU,
\qquad
U_+ + U_- = \frac{6 h_0^2}{\kappa} - \DelU \; .
\label{UfUt}
\end{equation}
Putting all the pieces together we obtain
\begin{subequations}
\begin{align}
B_0 
&=
\frac{27 \pi^2 S_1^4}{2 (\DelU)^3} 
=
\frac{27 \pi^2 (8\lambda^2)^4}{2 (\DelU)^3} \; ,
\label{B0o} \\
x 
&= 
\frac{3 S_1^2 \kappa \DelU}{4}
=
\frac{48 \lambda^4 \kappa}{\DelU},
\qquad
y = \frac{6 h_0^2}{\kappa \DelU} - 1 \; .
\label{xyo}
\end{align}
\end{subequations}
Moreover, we find
\begin{subequations}
\begin{align}
1 + 2xy + x^2
&=
(1 - x)^2 + 2 x \frac{6 h_0^2}{\kappa \DelU}
\nn\\
&=
\frac{1}{(\DelU)^2}
\left(
\left( \DelU - 48 \lambda^4\kappa \right)^2
+ 12 h_0^2 (48 \lambda^4) 
\right) ,
\label{1xyx2}
\end{align}
and
\begin{align}
1+xy
=
\frac{1}{(\DelU)^2}
\left(
(\DelU)^2 
-
48 \lambda^4 \kappa \DelU
+
6 h_0^2 (48 \lambda^4) 
\right).
\end{align}
\end{subequations}
Finally, we arrive at
\begin{align}
B &\simeq
\frac{27 \pi^2 (8\lambda^2)^4}{
\sqrt{\left( \DelU - 48 \lambda^4\kappa \right)^2 + 12 h_0^2 (48 \lambda^4)}
}
\nn \\
& \qquad \quad \times
\frac{1}{
(\DelU)^2 
-
48 \lambda^4 \kappa \DelU
+
6 h_0^2 (48 \lambda^4) 
+
\DelU
\sqrt{\left( \DelU - 48 \lambda^4\kappa \right)^2
+ 12 h_0^2 (48 \lambda^4)}
}
\nn\\
&=
\frac{2 \cdot 27 \pi^2 (8\lambda^2)^4 }{\sqrt{\left( \DelU - 48 
\lambda^4\kappa \right)^2
+ 12 h_0^2 (48 \lambda^4)}
} \nn \\
&\qquad \quad \times
\frac{1}{
\left(
\DelU
+
\sqrt{\left( \DelU - 48 \lambda^4\kappa \right)^2
+ 12 h_0^2 (48 \lambda^4)}
\right)^2
-
\left( 48 \lambda^4 \kappa
\right)^2
} \; ,
\label{BPoa}
\end{align}
which is the justification for \eqref{BPo} used in the main text.
%
%
\section{Instanton for sinusoidal potential and the thin-wall approximation}
\label{sec:inst}
%
In the thin-wall approximation \cite{Coleman:1980aw}, where the bubble radius 
$\bar{\rho}$ is much larger than the thickness of the surface wall of the 
bubble, one can neglect the change in $\rho$ at the wall. 
(A more quantitative definition of the thickness of the surface wall
is given below.)
The problem of finding an $O(4)$-symmetric 
bounce solution reduces to solving an instanton equation associated to the
following one-dimensional action:
\begin{subequations}
\begin{equation}
S_\psi
=
\int d\zeta
\left[
\frac{1}{2}
\dot{\psi}^2
+
U_0(\psi)
\right],
\label{Spsi}
\end{equation}
where the potential for the case of our interest is the one defined 
in \eqref{U0}, i.e.
\begin{equation}
U_0(\psi)
=
\lambda^4
\left(
1 - \cos \psi
\right) \;  .
\label{cospot}
\end{equation}
\end{subequations}
Note that $U_1(\psi)$ in \eqref{U1} differs from the 
constant part only by $\Ord(\DelU)$. 
Due to \eqref{TW1}, this difference is irrelevant in the equation of motion and 
can be dropped 
in the thin-wall approximation in the leading order.

The equation of motion derived from the action \eqref{Spsi} reads
\begin{equation}
\ddot{\psi} = \frac{\pa U_0}{\pa \psi} \; ,
\label{EoM}
\end{equation}
and can be solved as follows: Multiplying $\dot{\psi}$ on both sides 
of \eqref{EoM} and integrating once with respect to $\zeta$ gives
\begin{equation}
\frac{d \psi}{d\zeta}
=
\pm
\sqrt{2 U_0(\psi)}.
\label{insteq}
\end{equation}
The action of the solution $\psi_{\pm}$ of \eqref{insteq} becomes
\begin{align}
S_1 
\coloneqq
S_\psi [\psi_{\pm}]
=&
\int d\zeta
\left[
\frac{1}{2}
\dot{\psi}_{\pm}^2
+
U_0(\psi_{\pm})
\right]
=
\int d\zeta\,
2 U_0
=
\int d\psi\,
\sqrt{U_0}
\nn\\
=&
8 \lambda^2.
\label{S1a2}
\end{align}
The solution $\psi_\pm$ of the equation \eqref{insteq} is given by 
\begin{equation}
\psi_\pm (\zeta)
=
4 \tan^{-1}
\left[
\exp \left( \pm \lambda^2 (\zeta - \zeta_0) \right)
\right] \; ,
\label{instsol}
\end{equation}
where $\zeta_0$ is an integration constant.
The asymptotic behaviour of the instanton \eqref{instsol} at $\zeta \rightarrow 
+\infty$ is
\begin{equation}
\frac{\pi}{2}
-
\frac{\psi_+}{4}
\sim
\frac{\psi_-}{4}
\sim
e^{-\lambda^2(\zeta-\zeta_0)} .
\label{as+}
\end{equation}
The asymptotic behaviour of the instanton \eqref{instsol} at $\zeta \rightarrow 
-\infty$ is
\begin{equation}
\frac{\psi_+}{4}
\sim
\frac{\pi}{2}
-
\frac{\psi_-}{4}
\sim
e^{\lambda^2(\zeta-\zeta_0)} .
\label{as-}
\end{equation}
These asymptotic behaviours are anticipated from the mass term at the vacua.
These exponential decays define the thickness of the surface wall
to be $2/\lambda^2$.
Then the second condition \eqref{TW2} of the thin-wall approximate, applied to 
the bubble radius $\bar{\rho}$ defined in \eqref{barrho}, gives
\begin{equation}
\bar{\rho} \gg \frac{2}{\lambda^2}.
\label{TWconda}
\end{equation}
In the flat-space limit, we have
\begin{equation}
\bar{\rho} = \bar{\rho}_0
=
\frac{3 S_1}{\DelU}
=
\frac{24\lambda^2}{\DelU} \; .
\label{rho0ours}
\end{equation}
Inserting \eqref{rho0ours} into \eqref{TWconda}, we obtain the consistency 
condition for the thin-wall approximation in flat-space limit:
\begin{equation}
\frac{\DelU}{12\lambda^4} \ll 1 \;.
\label{fsTW}
\end{equation}
This condition is automatically satisfied due to the first condition 
\eqref{TW1} 
of the thin-wall approximation, applied to our model \eqref{TW1o}.
%
%
%
\bibliography{TunnellingDIref}

\providecommand{\href}[2]{#2}\begingroup\raggedright\begin{thebibliography}{10}

\bibitem{Freese:1990rb}
K.~Freese, J.~A. Frieman, and A.~V. Olinto, {\it {Natural inflation with pseudo
  - Nambu-Goldstone bosons}},  {\em Phys. Rev. Lett.} {\bf 65} (1990)
  3233--3236.

\bibitem{Banks:2010zn}
T.~Banks and N.~Seiberg, {\it {Symmetries and Strings in Field Theory and
  Gravity}},  {\em Phys. Rev.} {\bf D83} (2011) 084019,
  [\href{http://arxiv.org/abs/1011.5120}{{\tt arXiv:1011.5120}}].

\bibitem{ArkaniHamed:2003wu}
N.~Arkani-Hamed, H.-C. Cheng, P.~Creminelli, and L.~Randall, {\it {Extra
  natural inflation}},  {\em Phys.Rev.Lett.} {\bf 90} (2003) 221302,
  [\href{http://arxiv.org/abs/hep-th/0301218}{{\tt hep-th/0301218}}].

\bibitem{Banks:2003sx}
T.~Banks, M.~Dine, P.~J. Fox, and E.~Gorbatov, {\it {On the possibility of
  large axion decay constants}},  {\em JCAP} {\bf 0306} (2003) 001,
  [\href{http://arxiv.org/abs/hep-th/0303252}{{\tt hep-th/0303252}}].

\bibitem{ArkaniHamed:2006dz}
N.~Arkani-Hamed, L.~Motl, A.~Nicolis, and C.~Vafa, {\it {The String landscape,
  black holes and gravity as the weakest force}},  {\em JHEP} {\bf 06} (2007)
  060, [\href{http://arxiv.org/abs/hep-th/0601001}{{\tt hep-th/0601001}}].

\bibitem{Silverstein:2008sg}
E.~Silverstein and A.~Westphal, {\it {Monodromy in the CMB: Gravity Waves and
  String Inflation}},  {\em Phys. Rev.} {\bf D78} (2008) 106003,
  [\href{http://arxiv.org/abs/0803.3085}{{\tt arXiv:0803.3085}}].

\bibitem{McAllister:2008hb}
L.~McAllister, E.~Silverstein, and A.~Westphal, {\it {Gravity Waves and Linear
  Inflation from Axion Monodromy}},  {\em Phys. Rev.} {\bf D82} (2010) 046003,
  [\href{http://arxiv.org/abs/0808.0706}{{\tt arXiv:0808.0706}}].

\bibitem{Kaloper:2008fb}
N.~Kaloper and L.~Sorbo, {\it {A Natural Framework for Chaotic Inflation}},
  {\em Phys. Rev. Lett.} {\bf 102} (2009) 121301,
  [\href{http://arxiv.org/abs/0811.1989}{{\tt arXiv:0811.1989}}].

\bibitem{Berg:2009tg}
M.~Berg, E.~Pajer, and S.~Sjors, {\it {Dante's Inferno}},  {\em Phys. Rev.}
  {\bf D81} (2010) 103535, [\href{http://arxiv.org/abs/0912.1341}{{\tt
  arXiv:0912.1341}}].

\bibitem{Kaloper:2011jz}
N.~Kaloper, A.~Lawrence, and L.~Sorbo, {\it {An Ignoble Approach to Large Field
  Inflation}},  {\em JCAP} {\bf 1103} (2011) 023,
  [\href{http://arxiv.org/abs/1101.0026}{{\tt arXiv:1101.0026}}].

\bibitem{Franco:2014hsa}
S.~Franco, D.~Galloni, A.~Retolaza, and A.~Uranga, {\it {On axion monodromy
  inflation in warped throats}},  {\em JHEP} {\bf 02} (2015) 086,
  [\href{http://arxiv.org/abs/1405.7044}{{\tt arXiv:1405.7044}}].

\bibitem{Harigaya:2014rga}
K.~Harigaya and M.~Ibe, {\it {Phase Locked Inflation -- Effectively
  Trans-Planckian Natural Inflation}},  {\em JHEP} {\bf 11} (2014) 147,
  [\href{http://arxiv.org/abs/1407.4893}{{\tt arXiv:1407.4893}}].

\bibitem{Blumenhagen:2015kja}
R.~Blumenhagen, A.~Font, M.~Fuchs, D.~Herschmann, E.~Plauschinn, Y.~Sekiguchi,
  and F.~Wolf, {\it {A Flux-Scaling Scenario for High-Scale Moduli
  Stabilization in String Theory}},  {\em Nucl. Phys.} {\bf B897} (2015)
  500--554, [\href{http://arxiv.org/abs/1503.07634}{{\tt arXiv:1503.07634}}].

\bibitem{Ibanez:2015fcv}
L.~E. Ibanez, M.~Montero, A.~Uranga, and I.~Valenzuela, {\it {Relaxion
  Monodromy and the Weak Gravity Conjecture}},  {\em JHEP} {\bf 04} (2016) 020,
  [\href{http://arxiv.org/abs/1512.00025}{{\tt arXiv:1512.00025}}].

\bibitem{Hebecker:2015zss}
A.~Hebecker, F.~Rompineve, and A.~Westphal, {\it {Axion Monodromy and the Weak
  Gravity Conjecture}},  {\em JHEP} {\bf 04} (2016) 157,
  [\href{http://arxiv.org/abs/1512.03768}{{\tt arXiv:1512.03768}}].

\bibitem{Brown:2016nqt}
J.~Brown, W.~Cottrell, G.~Shiu, and P.~Soler, {\it {Tunneling in Axion
  Monodromy}},  {\em JHEP} {\bf 10} (2016) 025,
  [\href{http://arxiv.org/abs/1607.00037}{{\tt arXiv:1607.00037}}].

\bibitem{McAllister:2016vzi}
L.~McAllister, P.~Schwaller, G.~Servant, J.~Stout, and A.~Westphal, {\it
  {Runaway Relaxion Monodromy}},  \href{http://arxiv.org/abs/1610.05320}{{\tt
  arXiv:1610.05320}}.

\bibitem{Coleman:1977py}
S.~R. Coleman, {\it {The Fate of the False Vacuum. 1. Semiclassical Theory}},
  {\em Phys. Rev.} {\bf D15} (1977) 2929--2936. [Erratum: Phys.
  Rev.D16,1248(1977)].

\bibitem{Furuuchi:2014cwa}
K.~Furuuchi and Y.~Koyama, {\it {Large field inflation models from
  higher-dimensional gauge theories}},  {\em JCAP} {\bf 1502} (2015), no.~02
  031, [\href{http://arxiv.org/abs/1407.1951}{{\tt arXiv:1407.1951}}].

\bibitem{Furuuchi:2015jfj}
K.~Furuuchi and Y.~Koyama, {\it {The IR Obstruction to UV Completion for
  Dante's Inferno Model with Higher-Dimensional Gauge Theory Origin}},  {\em
  JCAP} {\bf 1606} (2016), no.~06 037,
  [\href{http://arxiv.org/abs/1511.06818}{{\tt arXiv:1511.06818}}].

\bibitem{Callan:1977pt}
C.~G. Callan, Jr. and S.~R. Coleman, {\it {The Fate of the False Vacuum. 2.
  First Quantum Corrections}},  {\em Phys. Rev.} {\bf D16} (1977) 1762--1768.

\bibitem{Baacke:2003uw}
J.~Baacke and G.~Lavrelashvili, {\it {One loop corrections to the metastable
  vacuum decay}},  {\em Phys. Rev.} {\bf D69} (2004) 025009,
  [\href{http://arxiv.org/abs/hep-th/0307202}{{\tt hep-th/0307202}}].

\bibitem{Dunne:2005rt}
G.~V. Dunne and H.~Min, {\it {Beyond the thin-wall approximation: Precise
  numerical computation of prefactors in false vacuum decay}},  {\em Phys.
  Rev.} {\bf D72} (2005) 125004,
  [\href{http://arxiv.org/abs/hep-th/0511156}{{\tt hep-th/0511156}}].

\bibitem{Coleman:1980aw}
S.~R. Coleman and F.~De~Luccia, {\it {Gravitational Effects on and of Vacuum
  Decay}},  {\em Phys. Rev.} {\bf D21} (1980) 3305.

\bibitem{Parke:1982pm}
S.~J. Parke, {\it {Gravity, the Decay of the False Vacuum and the New
  Inflationary Universe Scenario}},  {\em Phys. Lett.} {\bf B121} (1983)
  313--315.

\bibitem{Ade:2015lrj}
{\bf Planck} Collaboration, P.~A.~R. Ade et~al., {\it {Planck 2015 results. XX.
  Constraints on inflation}},  {\em Astron. Astrophys.} {\bf 594} (2016) A20,
  [\href{http://arxiv.org/abs/1502.02114}{{\tt arXiv:1502.02114}}].

\bibitem{Sasaki:1994yj}
M.~Sasaki, E.~D. Stewart, and T.~Tanaka, {\it {General solutions for tunneling
  of scalar fields with quartic potentials in de Sitter space}},  {\em Phys.
  Rev.} {\bf D50} (1994) 941--946,
  [\href{http://arxiv.org/abs/hep-ph/9402247}{{\tt hep-ph/9402247}}].

\end{thebibliography}\endgroup
\bibliographystyle{JHEP}
\end{document}